\documentclass[aps,showpacs,nofootinbib,eqsecnum,prd,twocolumn]{revtex4-1}
\usepackage{epsfig}  
\input{epsf}

\usepackage{graphicx}
\usepackage{amsmath}
\usepackage{amssymb}
\usepackage{marginnote}

\def\beq{\begin{equation}}
\def\eeq{\end{equation}}
\def\beqa{\begin{eqnarray}}
\def\eeqa{\end{eqnarray}}
\def\nn{\nonumber}

\begin{document}

\title{Decay constants of the pion and its excitations in holographic QCD}

\author{Alfonso Ballon-Bayona} \email{aballonb@fc.up.pt} 
\affiliation{Centro de F{\'i}sica do Porto,
Faculdade de Ci\^encias da Universidade do Porto, 
Rua do Campo Alegre 687, 4169-007 Porto, Portugal}

\author{Gast\~ao Krein} \email{gkrein@ift.unesp.br}
\affiliation{Instituto de F\'{\i}sica Te\'orica, Universidade Estadual
  Paulista,  Rua Dr. Bento Teobaldo Ferraz, 271 - Bloco II, 
  01140-070 S\~ao Paulo, SP, Brazil}

\author{Carlisson Miller} \email{miller@ift.unesp.br}
\affiliation{Instituto de F\'{\i}sica Te\'orica, Universidade Estadual
  Paulista,  Rua Dr. Bento Teobaldo Ferraz, 271 - Bloco II, 
  01140-070 S\~ao Paulo, SP, Brazil}


\begin{abstract}
We investigate the leptonic decay constants of the pion and its excitations with a 
5-d holographic model for quantum chromodynamics. We prove numerically that the leptonic 
decay constants of the excited states of the pion vanish in the chiral limit when chiral 
symmetry is dynamically broken. This nontrivial result is in agreement with a solid prediction 
of quantum chromodynamics and is based on a generalized Gell-Mann-Oakes-Renner relationship 
involving the decay constants and masses of the excited states of the pion. We also obtain 
an extended partially conserved axial-vector current relation that includes the fields of 
the excited states of the pion, a relation that was proposed long ago in the context of current algebra. 
\end{abstract}

\maketitle

\section{Introduction}

There is a solid prediction of quantum chromodynamics (QCD) that the leptonic
decay constant of the excited states of the pion vanish in the chiral limit when chiral symmetry 
is dynamically broken~\cite{Holl:2004fr}. The real world is not chirally symmetric,
as the masses of the $u$ and $d$ quarks are not zero. But these masses are much smaller 
than the strong-interaction scale $\Lambda_{\rm QCD}$ and it is therefore natural to expect 
that the leptonic decay constants of the excited states of the pion are dramatically suppressed in nature. 
At~first sight this prediction might seem surprising. Within a quark model perspective a 
suppression of the leptonic decay constants for excited states is expected; the 
leptonic decay constant for an $S$-wave state is proportional to the configuration-space 
wavefunction at the origin and, compared to the ground state, excited states have suppressed 
wavefunctions at the origin. However, within this perspective there is no obvious 
physical mechanism that suggests a dramatic reduction of the decay constants
for the excited states. The key point behind the suppression of the decay constants, 
as we shall elaborate shortly ahead, is the dynamical breaking of chiral symmetry 
in QCD and the (pseudo) Goldstone boson nature of the ground-state pion.

The suppression of the leptonic decay constants of pion's excited states is an interesting
feature of nonperturbative QCD. Lattice QCD and models of nonperturbative QCD can benefit 
from this feature by using it as a gauge to validate techniques and truncation schemes
in approximate calculations. A first lattice result from 2006~\cite{McNeile:2006qy} for the 
pion's first radial excitation, extrapolated to the chiral limit, gives $f_{\pi^1}/f_{\pi^0} 
\sim 0.08$~MeV; experimentally~\cite{Diehl:2001xe}, $f_{\pi^1}/f_{\pi^0} < 0.064$ -- the 
decay constant of the $n-$th excited state is denoted in the present paper by $f_{\pi^n}$ 
and that of the ground state by $f_{\pi^0}$. At about the same time, another lattice collaboration 
reports~\cite{Hashimoto:2008xg} a very small value for $f_{\pi^1}$, with an extrapolated value to
the chiral limit consistent with zero. Finally, a very recent 
publication~\cite{Mastropas:2014fsa} reports lattice results for the three lowest 
excited states: $f_{\pi^1}$ is modestly suppressed,$f_{\pi^2}$ is 
significantly suppressed, and $f_{\pi^3} \simeq f_{\pi^1}$. Calculations based on sum 
rules~\cite{{Kataev},{Elias:1997ya},{Maltman:2001gc}}, effective chiral 
Lagrangians~\cite{Volkov:1996br}, and a chiral quark model~\cite{Andrianov:1998kj} 
also find strongly suppressed values for $f_{\pi^1}$.  

In recent years a new class of models for tackling nonpertubative problems in QCD has received 
great attention in the literature. These are holographic models inspired on the gauge-gravity 
duality, in that a strongly coupled gauge theory in $d$ dimensions is assumed to be described 
equivalently in terms of a gravitational theory in $d + 1$ dimensions. The assumed duality is 
based on the Anti-de Sitter/Conformal Field Theory (AdS/CFT) 
correspondence~\cite{Maldacena,Gubser,Witten}, a conjectured relationship between conformal 
field theories and gravity theories in anti-de Sitter spaces -- for recent reviews, see 
Refs.~\cite{{Gubser:2009md},{CasalderreySolana:2011us}}. Although the holographic dual 
of QCD remains unknown, there exist several models attempting to construct the
five-dimensional holographic dual of QCD by incorporating known nonperturbative features
of QCD. Confinement, for example, can be modelled~\cite{Polchinski:2001tt} by truncating 
the AdS space with the introduction of an infrared cutoff $z_0 \sim 1/\Lambda_{\rm QCD}$ in the 
fifth dimension (the other four coordinates belong to the flat Minkowski spacetime).
In such a ``hard-wall'' model, one considers a slice $0 \leq z \leq z_0$ of AdS space, 
and imposes boundary conditions on the fields at the infrared border~$z_0$. Dynamical
chiral symmetry breaking can be incorporated~\cite{Erlich:2005qh,Da Rold:2005zs} in the 
hard-wall model with the use of scalar and vector fields in the AdS space which are
in correspondence, respectively, to the chiral order parameter and left- and right-handed 
currents of the $SU(2)_L \times SU(2)_R$ chiral flavor symmetry of QCD in Minkowski spacetime. 
In~this model the pion corresponds to the zero mode in the Kaluza-Klein expansion of the 
scalar field, it has a finite leptonic decay constant and its mass satisfies 
the Gell-Mann--Oakes--Renner (GOR) relationship. 

A particularly interesting approach in the holographic description of QCD is light-front 
holography (LFH), introduced in Refs.~\cite{{Brodsky:2006uqa},{Brodsky:2007hb}}.
In LFH, hadronic amplitudes in AdS space are mapped to frame-independent light-front 
wavefunctions in Minkowski space. This is made possible with the identification of the 
coordinate $z$  in AdS space with a Lorentz-invariant coordinate that measures 
the separation of the constituents within a hadron at equal light-front time. The way 
chiral symmetry is treated in LFH is nonstandard, as the vanishing of the pion mass in the 
chiral limit is not a result of the dynamical breaking of the symmetry, rather it follows 
from the precise cancellation of the light-front kinetic energy and light-front potential 
energy terms for the quadratic confinement potential in a Schr\"odinger-like 
equation~\cite{Brodsky:2013ar}. The same feature of obtaining a massless pion without DCSB 
is possible in a soft-wall LFH approach~\cite{Branz:2010ub}, in which confinement is modelled 
with a soft cutoff provided by a background dilaton field in the AdS space. The experimental 
values of the masses of the lowest radially and orbitally excited states of the pion are well 
reproduced, but the leptonic decay constants of the excited states do not vanish in the chiral limit. 

Motivated by these results in LFH, in the present paper we obtain the leptonic decay constants
of the pion and its excitations in a five-dimensional holographic hard wall model for QCD.
The decay constants are obtained directly from the Kaluza-Klein expansion of the holographic
currents, without resorting to LFH. We use the model of 
Ref.~\cite{Erlich:2005qh} and prove numerically that the leptonic decay constants of the 
excited states of the pion vanish in the chiral limit. In particular, we show that these
results follow from a generalized GOR relationship whose counterpart in QCD 
is~\cite{Holl:2004fr}
\begin{equation}
f_{\pi^n} \, m^2_{\pi^n} = 2 m_q \, \rho_{\pi^n},
\label{GOR-gen}
\end{equation}
where $m_{\pi^n}$ is the mass of the pion's $n$-th excited state, $m_q=m_u=m_d$ (we work in
the approximation of isospin symmetry) and $\rho_{\pi^n}$ is the gauge-invariant residue 
at the pole $P^2 = -m^2_{\pi^n}$ in the pseudoscalar vertex function; it is related to the 
matrix-valued Bethe-Salpeter wavefunction $\chi^a_{\pi^n}(P,q)$ via
\begin{equation}
i \rho_{\pi^n} \delta^{ab} := \int \frac{d^4 k}{(2\pi)^4} \, {\rm Tr} 
\left[t^a \gamma_5 \chi^b_{\pi^n}(q,P)\right],
\label{rho-def}
\end{equation}  
with the $SU(2)$ generators $t^a$, $a=1,2,3$, normalized as 
$2 \, {\rm Tr} \left( t^a t^b \right) = \delta^{ab}$. Although $m_q$ and $\rho_{\pi^n}$ are
scale dependent, the product $m_q \,\rho_{\pi^n}$ is renormalization group invariant. 
For the ground-state pion, DCSB implies~\cite{Maris:1997hd}  
\begin{equation}
\rho_{\pi^0} = - \frac{1}{f_{\pi^0}} \, \langle \bar q q \rangle,
\label{rho0}
\end{equation}
where $\langle \bar q q\rangle = \langle \bar u u \rangle = \langle \bar d d \rangle$ is 
the vacuum quark condensate; when this is used in Eq.~(\ref{GOR-gen}), the GOR relationship 
is obtained
\begin{equation}
f^2_{\pi^0} \, m^2_{\pi^0} = 2 m_q \, | \langle \bar q q\rangle|.
\label{GOR}
\end{equation}
As mentioned above, this key result of DCSB is obtained in holographic QCD in a rather 
straightforward way~\cite{Erlich:2005qh}. On the other hand, the vanishing
of the leptonic decay constants of the pion's excited states is more subtle
in holographic QCD, as we discuss in the present paper. In QCD, this key result follows
via the following chain of arguments~\cite{Holl:2004fr}: (1) the existence of excited states 
entails finite matrix-valued $\chi_{\pi^n}(P,q)$ wavefunctions; (2) the integral in 
Eq.~(\ref{rho-def}) is finite (this follows from the ultraviolet behavior of the QCD 
quark-antiquark scattering kernel); (3) then
\beq
\rho^0_{\pi^n} := \lim_{m_q \rightarrow 0} \rho_{\pi^n} = {\rm finite},
\label{rho-0}
\eeq
and (4) since, by hypothesis, $m^2_{\pi^n} \neq 0$ in the chiral limit, 
Eq.~(\ref{GOR-gen}) implies $f_{\pi^n} = 0$ for $m_q =0$. 

The paper is organised as follows. In the next section we present the hard wall model 
we use; we present the action and discuss how DCSB is implemented in the model, derive 
the equations of motion and holographic currents, and present the Kaluza-Klein expansion 
of the bulk fields. In section~\ref{sec:fpis} we derive the expressions for the leptonic 
decay constants via the Kaluza-Klein expansions for the holographic currents; we  
obtain an extended partially conserved axial-vector current relation that includes 
the fields of the excited states of the pion and derive the generalized GOR 
relationship (\ref{GOR-gen}). Numerical results are presented in section~\ref{sec:num}.
We discuss how the field equations are solved numerically, present results for the masses 
of the pion's ground and excited states, for the normalization of the field equations 
and for the function $\rho_{\pi^n}$ which appears in the generalized GOR relationship 
(\ref{GOR-gen}). Finally, we present  the results for the leptonic decay constants and
discuss the consistency of the results with the generalized GOR relationship.
Section~\ref{sec:concl} presents our conclusions and perspectives for future work. 

\section{The hard wall model}
\label{sec:model}

The AdS/QCD approach deals with the construction of 5-d holographic models for QCD-like 
theories by considering deformations of the AdS$_5$/CFT$_4$ correspondence. This bottom-up 
approach has proved to be very useful in describing many nonperturbative aspects of QCD 
and is complementary to the top-down approach, where 5-d holographic models arise as 
gravitational solutions of critical or non-critical string theories.

In the AdS/QCD approach, the simplest way to implement confinement is the so called 
hard-wall model, which consists of a slice of 5-d Anti-de-Sitter 
spacetime~\cite{Polchinski:2001tt}:
\beqa
ds^2 = \frac{1}{z^2} \left ( \eta_{\mu \nu}dx^{\mu}dx^{\nu} - dz^2 \right )\, ,   
\quad 0 < z  \le z_0 \,.
\label{AdSmetric}
\eeqa
where $\eta_{\mu \nu}=\text{diag} (1,-1,-1,-1)$ is the metric of 4-d flat 
spacetime. We are working in units where the AdS radius is unity.

The hard wall cut-off $z_0$ in the 5-d geometry corresponds to an infrared mass gap in the 4-d 
gauge theory, $z_0 = 1/\Lambda_{\rm QCD}$. As a consequence, conformal symmetry is broken and the 
theory is confining, as can be shown through the holographic calculation of the quark 
anti-quark potential~\cite{Kinar:1998vq}.

\subsection{DCSB implementation in holographic QCD}

DCSB was first implemented in  holographic QCD in Refs.~\cite{Erlich:2005qh,Da Rold:2005zs}. 
Here we will follow the conventions and notation
of Ref.~\cite{Erlich:2005qh}. The AdS/CFT correspondence maps 4-d field theory operators
$\mathcal{O}(x)$ to 5-d fields $\phi\left(x,z\right)$. In the case of QCD the relevant operators
for describing DCSB are the left and right handed currents 
$J_{L\mu}^{a}=\bar{q}_{L}\gamma_{\mu}t^{a}q_{L}$, $J_{R\mu}^{a}=\bar{q}_{R}\gamma_{\mu}t^{a}q_{R}$,
corresponding to the $SU(N_{f})_{L}\times SU(N_{f})_{R}$ chiral flavor symmetry and the quark 
bilinear operator $\bar{q}_{R}{q}_{L}$ related to DCSB. In the dual theory, these 4-d operators 
correspond to 5-d gauge fields $L_{m}^{a}(x,z)$, $R_{m}^{a}(x,z)$ and a 5-d bifundamental scalar 
field $X(x,z)$, both living in an AdS slice described by Eq.~(\ref{AdSmetric}).  

Since the mass of a p-form in 5-d AdS spacetime is related to the  dimension $\Delta$ of the 
dual 4-d operator via the relation $m^{2}=\left(\Delta-p\right)\left(\Delta+p-4\right)$, 
one has that the gauge fields $L_{m}^{a}(x,z)$, $R_{m}^{a}(x,z)$ are massless whereas the 
scalar field $X(x,z)$ has a negative mass squared $m^2=-3$.   

The action in Ref.~\cite{Erlich:2005qh} can be written as 
\beqa
S &=& \int d^5 x \sqrt{|g|} \, {\rm Tr}\biggl[ (D^m X)^{\dag} (D_m X)  + 3 |X|^2
\nn \\
&& - \frac{1}{4 g_5^2} \left ( L^{mn} L_{mn} + R^{mn} R_{mn}  \right )
\biggr] ,  
\label{Erlichaction}
\eeqa
where 
\beqa
D_m X &:=& \partial_m X- i L_m X + i X R_m  , \label{X} \\[0.2true cm]
L_{mn} &:=& \partial_m L_n - \partial_n L_m - i \left [ L_m , L_n \right ] , 
\label{Lmn} \\[0.2true cm]
R_{mn} &:=& \partial_m R_n - \partial_n R_m - i \left [ R_m , R_n \right ] .
\label{Rmn}
\eeqa
The action includes the $N_f$ gauge fields $L_m$ and $R_m$, corresponding to the 
left and right flavor currents in QCD, and the  bifundamental scalar $X$ dual to the quark 
bilinear operator $\bar q_R q_L$. In this paper we restrict the discussion to the case 
$N_f=2$, corresponding to the quark flavors $u$ and $d$. The dynamics of the  5-d fields $L_m$, $R_m$ and $X$ is 
described by the action in Eq.~(\ref{Erlichaction}) and the classical solution that describes chiral
symmetry breaking is given by  
\beqa
L^0_m = & R^0_m = 0 \, , \quad 2 X_0 = \zeta M z + \frac{\Sigma}{\zeta} z^3 ,
\label{classicalfields}
\eeqa
with
\beqa
M =
 \begin{pmatrix}
  m_u & 0 \\
  0 & m_d
 \end{pmatrix} , \quad
 \Sigma =
 \begin{pmatrix}
  \sigma_u & 0 \\
  0 & \sigma_d
 \end{pmatrix} .
\eeqa
The parameter $\zeta=\sqrt{N_c}/(2\pi)$ in Eq.~(\ref{classicalfields}) is introduced
to be consistent with the counting rules of large-$N_c$ QCD~\cite{Cherman:2009}. 
The AdS/CFT dictionary leads to identification of the coefficients $M$ and $\Sigma$ 
with the 4-d quark mass and chiral condensate terms responsible for the explicit and 
dynamical breaking of chiral symmetry. 

To investigate the consequences of DCSB on the mesons we consider 
perturbations around the background fields in Eq.~(\ref{classicalfields}). First of all 
it is convenient to rewrite the gauge field fluctuations in terms of vectorial and axial 
fields
\beqa
L_m = V_m + A_m  \quad , \quad
R_m = V_m - A_m . \label{VApert}
\eeqa
The bifundamental field $X$ is decomposed into the classical part $X_0$ and a 
pseudoscalar fluctuation $\pi$ in the following form~\cite{Abidin:2009aj}:
\beqa
X =  e^{ i  \pi^a t^a } X_0 \, e^{i\pi^a t^a } .
\label{Xexp}
\eeqa
We will work in the isospin symmetrical limit, $m_u = m_d =: m_q$ 
and $\sigma_u = \sigma_d =: \sigma_d$;  in this limit the matrix  $X_0$ is 
proportional to the unit matrix and $X$ becomes
\beq
X = X_0 \, e^{2 i\pi^a t^a}.
\eeq 

The meson spectrum is obtained from  the kinetic terms; we expand the original action 
(\ref{Erlichaction}) up to quadratic order in $V_m = V^a_m \, t^a$, 
$A_m = A^a \, t^a$ and the fluctuation $\pi^a$:  
\beqa
S^{\rm Kin } &=& \int d^5 x \sqrt{|g|} \,
\biggl[   \frac{v^2}{2} (\partial_m \pi^a  - A_m^a)^2 \nn \\
&& - \frac{1}{4 g_5^2} (v^{mn}_a v_{mn}^a
+ a^{mn}_a a_{mn}^a ) \biggr] ,
\label{kineticterm}
\eeqa
with 
\beq
v_{mn}^a := \partial_m  V_n^a - \partial_n V_m^a , \quad
a_{mn}^a := \partial_m  A_n^a - \partial_n A_m^a ,
\label{vmn&amn}
\eeq
and 
\beqa
v(z) := \zeta m_q z + \frac{\sigma_q}{\zeta}  z^3 .
\label{v}
\eeqa

\subsection{Equations of motion and holographic currents}

Writing $S^{\rm Kin}$ as
\beq
S^{\rm Kin} = \int d^5 x \; {\cal L}^{\rm Kin},
\eeq
its variation takes the Euler-Lagrange form
\beqa
\delta S^{\rm Kin} &=& \int d^5 x \, \Biggl [ \left ( \frac{\partial {\cal L}^{\rm Kin}}
{\partial V_\ell^a} - \partial_m P_{V , a} ^{m \ell}\right ) \, \delta V_{\ell}^a
\nn \\
&& \hspace{-1.2cm} + \left ( \frac{\partial {\cal L}^{\rm Kin}}{\partial A_\ell^a}
- \partial_m P_{A , a} ^{m \ell}\right ) \, \delta A_{\ell}^a 
+ \left ( \frac{\partial {\cal L}^{\rm Kin}}{\partial \pi^a}
- \partial_m P_{\pi , a}^m \right ) \, \delta \pi^a \Biggr] \nn \\
&& \hspace{-1.2cm} + \int d^5 x \, \partial_m \left( P_{V , a}^{m \ell} \;
\delta V_{\ell}^a + P_{A , a}^{m \ell} \, \delta A_{\ell}^a
+ P_{\pi , a}^m \, \delta \pi^a \right) , 
\label{ActionVariation}
\eeqa
where $P_{V , a} ^{m \ell}$ and  $P_{A , a} ^{m \ell}$ are the conjugate momenta
to the vector fields $V_\ell^a$ and $A_\ell^a$
\beq
P_{V , a} ^{m \ell} :=
\frac{\partial {\cal L}^{\rm Kin}}{ \partial (\partial_m V_\ell^a) }, 
\hspace{1.0cm}
P_{A , a} ^{m \ell} :=
\frac{\partial {\cal L}^{\rm Kin}}{ \partial (\partial_m A_\ell^a) },
\eeq
and $P_{\pi , a}^m$ the conjugate momentum to the field $\pi^a$
\beq
P_{\pi , a}^m :=
\frac{\partial {\cal L}^{\rm Kin}}{\partial ( \partial_m \pi^a ) } .
\eeq
From Eq.~(\ref{kineticterm}) we find for the derivatives of ${\cal L}^{\rm Kin}$
with respect to the fields
\beqa
&& \frac{\partial {\cal L}^{\rm Kin}}{\partial V_\ell^a}  = 
\frac{\partial {\cal L}^{\rm Kin}}{\partial \pi^a} = 0 , \\
&& \frac{\partial {\cal L}^{\rm Kin}}{\partial A_\ell^a}  = - v^2 \sqrt{|g|}  
\left ( \partial^\ell \pi_a - A^\ell_a \right ) ,
\eeqa
and for the conjugate momenta
\beqa
&& P_{V , a} ^{m \ell} = - \frac{1}{g_5^2} \sqrt{|g|} v^{m \ell}_a , \\
&& P_{A , a} ^{m \ell} = - \frac{1}{g_5^2} \sqrt{|g|} a^{m \ell}_a , \\
&& P_{\pi , a}^m = v^2 \sqrt{|g|}  \left ( \partial^m \pi_a - A^m_a \right ) .
\eeqa

Imposing the stationarity condition $\delta S^{\rm Kin} = 0$, we find from the 
first three terms in Eq.~(\ref{ActionVariation}) the field equations 
\beqa
&& \hspace{-0.5cm}\partial_m \left ( \sqrt{|g|} \, v^{mn}_a \right ) = 0 , 
\label{Fieldeq1} \\[0.2true cm]
&& \hspace{-0.5cm}\partial_m \left ( \sqrt{|g|} \, a^{mn}_a \right )
- g_5^2 v^2 \sqrt{|g|} \left ( \partial^n \pi^a - A^n_a \right ) = 0 , 
\label{Fieldeq2} \\[0.2true cm]
&& \hspace{-0.5cm}\partial_m \left [ v^2 \sqrt{|g|}  \left(\partial^m \pi^a 
- A^m_a \right) \right ] = 0 .
\label{Fieldeq3}
\eeqa
The mass spectrum of the vector and axial-vector mesons and pions can be found
by solving these equations of motion in momentum space under appropriate boundary 
conditions. We will choose, however, a different method based on the Kaluza-Klein 
expansion. As explained in the next subsection, the advantage of the Kaluza-Klein 
method is the extraction of a 4-d off-shell action for the mesons. 

The last three terms in (\ref{ActionVariation}) form  a surface term,  
whose nonvanishing contribution can be written as
\beqa
\delta S^{\rm Kin}_{\rm Bdy} &=& \int d^4 x  \Bigl( P_{V , a}^{z \mu} \, \delta V_{\mu}^a
+ P_{A , a}^{z \mu} \, \delta A_{\mu}^a \nn \\[0.2true cm]
&& +\,  P_{\pi , a}^z \, \delta \pi^a \Bigr)^{z=z_0}_{z=\epsilon}.
\eeqa
The terms at $z=z_0$ vanish under Neummann boundary conditions
\beqa
\partial_z V_\mu^a |_{z=z_0} = \partial_z A_\mu^a |_{z=z_0} 
= \partial_z \pi^a |_{z=z_0} = 0 ,
\eeqa
and for the gauge choice $V_z^a = A_z^a = 0$. The boundary terms at  
$z=\epsilon$ can be written as (we distinguish vectorial Minkowski indices 
$\hat \mu$ and vectorial AdS indices $\mu$) 
\beqa
\delta S^{\rm Kin}_{\rm Bdy} &=& - \int d^4 x  \,  \Bigl [ \langle J^{\hat \mu}_{V,a} \rangle 
\, \left( \delta V_{\hat \mu}^a \right)_{z = \epsilon}
+  \langle J^{\hat \mu}_{A,a} \rangle  \, \left(\delta A_{\hat\mu}^a\right)_{z = \epsilon}
\nn \\[0.2true cm]
&& +  \, \langle J_{\pi,a} \rangle \, \left(\delta \pi^a \right)_{z = \epsilon} \Bigr ] ,
\eeqa
where we find the holographic currents 
\beqa
\langle J^{\hat\mu}_{V,a} (x) \rangle &=&  P_{V , a}^{z \mu} \vert_{z = \epsilon}
= - \frac{1}{g_5^2} \left ( \sqrt{|g|} v^{z \mu} \right )_{z = \epsilon} , 
\label{HologCurrent1} \\
\langle J^{\hat\mu}_{A,a} (x) \rangle &=&  P_{A , a}^{z \mu} \vert_{z = \epsilon}
= - \frac{1}{g_5^2} \left ( \sqrt{|g|} a^{z \mu} \right )_{z = \epsilon}  , 
\label{HologCurrent2} \\
\langle J_{\pi,a} (x) \rangle &=&
 P_{\pi , a}^z \vert_{z = \epsilon}
=   \left [ \sqrt{|g|} v^2 \left ( \partial^z \pi_a - A^z_a \right ) \right ]_{z = \epsilon}
\nn \\[0.2true cm]
&=& \partial_{\mu} \langle J^{\mu}_{A,a} (x) \rangle . 
\label{HologCurrent3}
\eeqa
These holographic currents are identified with the vacuum expectation values 
of the 4-d vectorial, axial and pion current operators in Minkowski spacetime.

\subsection{The Kaluza-Klein expansion}

First of all we evaluate the metric (\ref{AdSmetric}) to write the action 
of Eq.~(\ref{kineticterm}) in the form
\beqa
&&\hspace{-0.75cm}
S^{\rm Kin} =  \int d^4 x \int \frac{dz}{z} \,  \biggl\{\frac{v^2}{2 z^2}
\Bigl[ - \left(\partial_z \pi^a - A_z^a\right)^2  \nn \\
&&\hspace{-0.5cm}  + \, \left(\partial_{\hat \mu} \pi^a - A_{\hat \mu}^a\right)^2 \Bigr] \nn \\
&&\hspace{-0.5cm}  - \frac{1}{4 g_5^2} \Bigl[  - 2 (v_{z \hat \mu}^a)^2 + (v_{\hat \mu \hat \nu}^a)^2
- \,  2 (a_{z \hat \mu}^a)^2 + (a_{\hat \mu \hat \nu}^a)^2  \Bigr ]
\biggr\} . \label{kinetictermv2}
\eeqa
The axial-vector  field $A_{\hat \mu}^a$ can be decomposed into transverse and longitudinal 
parts:
\beqa
A_{\hat \mu}^a = A_{\hat \mu}^{\perp ,a} + \partial_{\hat \mu} \phi^a \,,
\eeqa
where $\partial_{\hat \mu} A^{\hat \mu}_{\perp ,a} = 0$. The transverse part will be 
associated with the axial-vector mesons whereas the longitudinal part will be associated 
with the pions. The  action (\ref{kinetictermv2}) is invariant under the gauge
transformations
\begin{eqnarray}
V^{a}_{m} &\rightarrow& V^{a}_{m}-\partial_{m}\lambda_V^{a}, \\[0.2true cm]
A^{a}_{m} &\rightarrow& A^{a}_{m}-\partial_{m}\lambda_A^{a}, \\[0.2true cm]
 \pi^{a} &\rightarrow& \pi^{a} - \lambda_A^{a}.
\end{eqnarray}
Because of this, we can fix the gauge as $V_z^a = A_z^a = 0$ and  then 
Eq.~(\ref{kinetictermv2}) reduces to
\beqa
S^{\rm Kin} &=&  \int d^4 x \int \frac{dz}{z} \, \Big \{ \frac{v^2}{2 z^2}
\Bigl [ - \left(\partial_z \pi^a \right)^2 
+ \left(\partial_{\hat \mu} \pi^a - \partial_{\hat \mu} \phi^a\right)^2 \nn \\
&& + \, \left(A_{\hat \mu}^{\perp ,a}\right)^2  \Bigr] 
+ \partial_{\hat \mu} (\cdots)  \nn \\
&& - \frac{1}{4 g_5^2} \Big [  - 2 \left(\partial_z V_{\hat \mu}^a\right)^2 
+ \left(v_{\hat \mu \hat \nu}^a\right)^2 - 2 \left(\partial_z A_{\hat \mu}^a\right)^2  
\nn \\
&& - 2 \left(\partial_z \partial_{\hat \mu} \phi^a\right)^2 
+ \left(a_{\hat \mu \hat \nu}^{\perp ,a}\right)^2  \Big ]  \Big \} \,, \label{kinetictermv3}
\eeqa
where the terms in $(\cdots)$ vanish by choosing appropriate boundary conditions.

The action (\ref{kinetictermv3}) is in a suitable form for the Kaluza-Klein expansion. 
This consists in expanding the 5-d fields in an infinite discrete set of modes. Each mode 
will be a product of a pure wave function in the radial coordinate $z$ and a meson field 
depending on the Minkowski coordinates $x$. For the present case, the Kaluza-Klein expansion 
for the bulk fields $V_{\hat \mu}^a$, $A_{\hat \mu}^{\perp,a}$, $\pi^a$ and $\phi^a$ take the 
form 
\beqa
V_{\hat \mu}^a (x,z) &=& g_5 \sum_{n=0}^{\infty} v^{a,n}(z) \hat V_{\hat \mu}^{a,n} (x) ,
\label{KKexpansion1} \\
A_{\hat \mu}^{\perp,a} (x,z) &=& g_5 \sum_{n=0}^{\infty} a^{a,n}(z) \hat A_{\hat \mu}^{a,n} (x) , 
\label{KKexpansion2} \\
\pi^a(x,z) &=& g_5 \sum_{n=0}^{\infty} \pi^{a,n}(z) \hat \pi^{a,n}(x), 
\label{KKexpansion3} \\
\phi^a(x,z) &=& g_5 \sum_{n=0}^{\infty} \phi^{a,n}(z) \hat \pi^{a,n}(x) .
\label{KKexpansion4}
\eeqa
The wave functions $\phi^{a,n}(z)$ and $\pi^{a,n}(z)$ are not independent. The relation 
between them can be obtained from the 5-d field equations in 
Eqs.~(\ref{Fieldeq1})-(\ref{Fieldeq3}) or via an off-shell integration, as described below.

Using  the Kaluza-Klein expansions (\ref{KKexpansion1})-(\ref{KKexpansion4}) in the 
action (\ref{kinetictermv3}), one can separate the $z$ and $x$ integrals and write the 
action as 
\beqa
S^{\rm Kin} &=&  \sum_{n,m=0}^{\infty} \int d^4 x \, \biggl \{
\frac12 \Delta_{\pi}^{a,nm} \, \left[\partial_{\hat \mu} \pi^{a,n} (x)\right]
\partial^{\hat \mu} \pi^{a,m} (x) \nn \\
&& \hspace{-1.0cm} - \frac12 M_{\pi}^{a,nm}   \, \hat \pi^{a,n}(x) \, \hat \pi^{a,m}(x) 
- \frac{1}{4} \Delta_{V}^{a,nm} \, \hat v_{\hat \mu \hat \nu}^{a,n}(x) \, 
\hat v^{\hat \mu \hat \nu}_{a,m}(x)\nn \\
&& \hspace{-1.0cm} 
+ \frac{1}{2} M_V^{a,nm} \, \hat V_{\hat \mu}^{a,n}(x) \, \hat V^{\hat \mu}_{a,m}(x) 
- \frac{1}{4} \Delta_{A}^{a,nm} \, \hat a_{\hat \mu \hat \nu}^{a,n}(x) \, 
\hat a^{\hat \mu \hat \nu}_{a,m}(x)  \nn \\
&& \hspace{-1.0cm}
+ \frac12 M_A^{a,nm} \, \hat A_{\hat \mu}^{a,n}(x) \, \hat A^{\hat \mu}_{a,m}(x) \biggr \}, \label{kinetictermv4}
\eeqa
where the coefficients for the 4-d fields are given by the following $z$ integrals
\beqa
&&\hspace{-0.5cm}\Delta_{\pi}^{a,nm} =  \int \frac{dz}{z} \, \Bigl\{
\left[\partial_z \phi^{a,n}(z)\right] \partial_z \phi^{a,m}(z) \nn \\
&&\hspace{-0.25cm} + \, \beta(z)  
\left[\pi^{a,n}(z) - \phi^{a,n}(z)\right] 
\left[\pi^{a,m}(z) - \phi^{a,m}(z)\right] 
\Bigr\} , \label{Delta-pi} \\
%
&&\hspace{-0.5cm}\Delta_{V}^{a,nm} =  \int \frac{dz}{z} \, v^{a,n}(z) \, v^{a,m}(z) , \nn \\
&&\hspace{-0.5cm}\Delta_{A}^{a,nm} = \int \frac{dz}{z} \,  a^{a,n}(z) \, a^{a,m}(z)  , \\
%
&&\hspace{-0.5cm}M_{\pi}^{a,nm} =  \int \frac{dz}{z} \, \beta(z) \left[\partial_z \pi^{a,n}(z)\right] 
\partial_z \pi^{a,m}(z) , \nn \\
&&\hspace{-0.5cm}M_V^{a,nm} =  \int \frac{dz}{z} \, \left[\partial_z v^{a,n}(z)\right] 
\partial_z v^{a,m}(z) , \nn \\ 
&&\hspace{-0.5cm}M_A^{a,nm} =  \int \frac{dz}{z} \, \Bigl\{  
\left[\partial_z a^{a,n}(z)\right] \partial_z a^{a,m}(z) \nn \\
&& + \, \beta(z) a^{a,n}(z) a^{a,m}(z)\Bigr\} ,
\eeqa
and we have defined $\beta(z)$ as
\beqa
\beta (z) := \frac{g_5^2}{z^2}  {v(z)}^2 = g_5^2 \left ( \zeta m_q + 
\frac{\sigma_q}{\zeta} z^2 \right )^2\,. \label{defbeta}
\eeqa

Now the goal is to obtain a 4-d action with standard kinetic terms for the vector 
$\hat{V}_{\hat \mu}^{a,n}$ and axial-vector ${\hat A}^{a,n}_{\hat \mu}$ mesons 
and pions $\pi^{a,n}$. This can be achieved imposing the conditions
\beqa
\Delta_{\pi}^{a,nm} &=& \Delta_{V}^{a,nm} = \Delta_{A}^{a,nm} = \delta^{nm}  \, ,  
\label{Deltacond} \\[0.2true cm]
M_{\pi}^{a,nm} &=&  m_{\pi^{a,n}}^2 \delta^{nm}, \, \,   \, \,
M_{V}^{a,nm} =  m_{V^{a,n}}^2 \delta^{nm} , \label{masscond1}
\\[0.2true cm]
M_{A}^{a,nm} &=&  m_{A^{a,n}}^2 \delta^{nm} . 
\label{masscond2}
\eeqa
This way we arrive at the following 4-d action :
\beqa
S^{\rm Kin} &=&  \sum_{n=0}^{\infty}  \int d^4 x \, \biggl \{
\frac12 \left[\partial_{\hat \mu} \hat \pi^{a,n} (x)\right]^2 
- \frac12 m_{\pi^{a,n}}^2   \left[\hat \pi^{a,n}(x)\right]^2 \nn \\
&& - \frac{1}{4} \left[\hat v_{\hat \mu \hat \nu}^{a,n}(x)\right]^2
+ \frac{1}{2} m_{V^{a,n}}^2 \left[\hat V_{\hat \mu}^{a,n}(x)\right]^2 
\nn \\ 
&& \ - \frac{1}{4} \left[\hat a_{\hat \mu \hat \nu}^{a,n}(x)\right]^2 
+ \frac12 m_{A^{a,n}}^2 \left[\hat A_{\hat \mu}^{a,n}(x)\right]^2 \biggr \} . 
\label{kinetictermv5}
\eeqa
The conditions (\ref{Deltacond}) are precisely the normalization rules for 
the wave functions $v^{a,n}(z)$, $a^{a,n}(z)$, $\pi^{a,n}(z)$ and $\phi^{a,n}(z)$.
Moreover, the conditions on the masses, Eqs.~(\ref{masscond1}) and (\ref{masscond2}),
can be obtained from the normalization conditions (\ref{Deltacond}) as long 
as we impose the following equations 
\beqa
&& \hspace{-0.6cm}\frac{\beta(z)}{z} \left[\pi^{a,n}(z) - \phi^{a,n}(z) \right] = 
- \partial_z \left [\frac{1}{z} \partial_z \phi^{a,n}(z) \right ] ,
\label{EQ1} \\[0.2true cm]
&& \hspace{-0.6cm}\beta(z) \, \partial_z \pi^{a,n}(z) = m_{\pi^{a,n}}^2 \partial_z 
\phi^{a,n}(z) , 
\label{EQ2} \\[0.2true cm]
&& \hspace{-0.6cm} - \partial_z \left [ \frac{1}{z} \partial_z   v^{a,n}(z) \right ] = \frac{m_{V^{a,n}}^2}{z} 
v^{a,n}(z) \, , \label{EQ3} \\[0.2true cm]
&& \hspace{-0.6cm} \left [ - \partial_z \left ( \frac{1}{z} \partial_z   \right )
+  \frac{1}{z} \beta(z) \right ]  a^{a,n}(z) = \frac{m_{A^{a,n}}^2}{z} a^{a,n}(z)  
\label{EQ4} .
\eeqa
These equations can be interpreted as the on-shell conditions for the wave functions 
$v^{a,n}(z)$, $a^{a,n}(z)$, $\pi^{a,n}(z)$ and $\phi^{a,n}(z)$. It should be noted
that these equations can simply be obtained from the 5-d field equations in
Eqs.~(\ref{Fieldeq1})-(\ref{Fieldeq3}). However, the convenience of our method is 
the fact that the final 4-d action (\ref{kinetictermv5}) remains off-shell, a property 
that can be very useful when considering scattering amplitudes, which require the 
knowledge of Feynman rules. 

To conclude, we mention the meson masses $m_{\pi^{a,n}}$, $m_{V^{a,n}}$ and $m_{A^{a,n}}$ 
are completely determined once we find solutions for Eqs.~(\ref{EQ1})-(\ref{EQ4}). In the
following section, solutions to these equations are found imposing Dirichlet boundary 
conditions at $z=\epsilon$:
\beqa
\pi^{a,n} \vert_{z=\epsilon}  = v^{a,n}  \vert_{z=\epsilon}  
= a^{a,n} \vert_{z=\epsilon}  = 0,
\label{DirichletBC}  
\eeqa
and Neumann boundary conditions at $z=z_0$:
\beqa
\hspace{-0.5cm}
\partial_z \pi^{a,n} \vert_{z=z_0}  = \partial_z v^{a,n} \vert_{z=z_0} 
= \partial_z a^{a,n} \vert_{z=z_0} = 0  . 
\label{NeumannBC}
\eeqa
Since there is no flavor mixing, the equations of motion and boundary conditions 
have the same form for every flavor index $a$ and from now on we will omit that index. 
The latter is  relevant for the calculation of pion decay constants.

\section{Leptonic decay constants of the excited states of the pion}
\label{sec:fpis}

In this section we present the main result of our paper; the behavior of the leptonic
decay constants of the exited states of the pion near the chiral limit. First we show how 
to calculate the decay constants $f_{\pi^n}$ from holography and arrive at the equivalent 
to Eq.~(\ref{GOR-gen}) in QCD. After describing some technical details on the calculation of
the decay constants, we present our numerical results for the~$f_{\pi^n}$. 

\subsection{Holographic calculation of the decay constants}
\label{sub:holfpi}

The simplest way to extract the leptonic decay constant is to use 
the Kaluza-Klein expansions (\ref{KKexpansion1})-(\ref{KKexpansion4}) 
for the holographic currents (\ref{HologCurrent1})-(\ref{HologCurrent3}):
\beqa
\langle J^{\hat \mu}_{V}  (x) \rangle &=&   \sum_{n=0}^{\infty} 
\left [  \frac{1}{g_5 z} \partial_z v^{n}(z)  \right ]_{z=\epsilon} \hat V^{\hat \mu}_n(x) ,
\label{JVExp}\\
\langle J^{\hat \mu}_{A}  (x) \rangle &=&   \sum_{n=0}^{\infty} \left [  \frac{1}{g_5 z } 
\partial_z a^{n}(z)   \right ]_{z=\epsilon} \hat A^{\hat \mu}_{n}(x) \nn \\
&+& \sum_{n=0}^{\infty} \left [ \frac{1}{g_5 z} \partial_z \phi^{n}(z) \right ]_{z=\epsilon}
\partial^{\hat \mu} \hat \pi^{n}(x) , \label{JAExp} \\
\partial_{\hat \mu} \langle J^{\hat \mu}_{A} (x) \rangle &=& \langle J_{\Pi}  (x) \rangle \cr
&=&   - \sum_{n=0}^{\infty} \left [  \frac{ \beta (z) }{g_5 z}\partial_z \pi^{n}(z)  
\right ]_{z=\epsilon} \hat \pi_{n}(x) .  
\label{JPiExp}
\eeqa
As explained in the previous section, we are omitting the flavor index $a$. Here the 4-d 
fields $\hat V^{\hat \mu}_{n}(x)$, $\hat A^{\hat \mu}_{n}(x)$ and $\hat \pi^{n}(x)$ are 
on-shell. From the current expansions (\ref{JVExp}) and (\ref{JAExp}) we are able to extract 
the decay constants for the vector mesons ($g_{V^{n}}$), the axial-vector mesons ($g_{A^{n}}$), 
and the pions  ($f_{\pi^{n}}$):
\beqa
g_{V^{n}} &=& \left [  \frac{1}{g_5 z} \partial_z v^{n}(z)  \right ]_{z=\epsilon}, 
\label{gV}
\\
g_{A^{n}} &=& \left [  \frac{1}{g_5 z} \partial_z a^{n}(z)  \right ]_{z=\epsilon} , 
\label{gA}
\\
f_{\pi^{n}} &=& \left [ - \frac{1}{g_5 z} \partial_z \phi^{n}(z)  \right ]_{z=\epsilon} .
\label{fpin}
\eeqa
These results are consistent with the standard definition of meson decay constants: 
\beqa
&& \langle 0 \vert J^{\hat \mu}_V (0) \vert V_n (p,\lambda) \rangle =
    \epsilon^{\hat \mu}(p,\lambda) g_{V^n}, \\
&& \langle 0 \vert J^{\hat \mu}_A (0) \vert A_n (p,\lambda) \rangle = 
   \epsilon^{\hat \mu}(p,\lambda) g_{A^n} , \\
&&
\langle 0 \vert J^{\hat \mu}_A (0) \vert \pi_n (p) \rangle = 
 i p^{\hat \mu} \,  f_{\pi^n}  .
\eeqa

Taking the divergence of (\ref{JAExp}) and using (\ref{fpin}), one obtains the interesting relation
\beqa
f_{\pi^{n}} m_{\pi^{n}}^2 = - \frac{1}{g_5} \left [  \frac{\beta (z)}{ z} \partial_z \pi^{n}(z)  
\right ]_{z=\epsilon} ,
\label{HGOR}
\eeqa
where we made use of the on-shell equation for the pion field,  
$\partial^2 \hat \pi^{n}(x) = -m_{\pi^{n}}^2 \hat \pi^{n}(x)$. Moreover, using this result into 
Eq.~(\ref{JPiExp}), the divergence of the axial current takes the form of an extended PCAC 
relation
\beqa
\partial_\mu \langle J^{\hat \mu}_{A}  (x) \rangle 
= \sum_{n=0}^{\infty}  f_{\pi^{n}} m_{\pi^n}^2 \hat \pi_{n}(x) .
\label{ext-PCAC}
\eeqa
Interestingly, such a relation was proposed long ago~\cite{Dominguez:1976ut} in the
context of current algebra studies. In particular, such an extended PCAC relation 
leads naturally to the vanishing of the leptonic decay constants of pion's excited 
states when the ground-state pion is the Goldstone boson of DCSB and $m_{\pi^n} \neq 0$, 
$n \ge 1$, in the chiral limit. 
Moreover, from Eq.~(\ref{HGOR}) one obtains a generalized GOR relationship in the form 
of Eq.~(\ref{GOR-gen}) if one makes the identification
\beqa
2 m_q  \rho_{\pi^{n}} := - \frac{1}{g_5} \left [  \frac{\beta (z)}{ z} \partial_z \pi^{n}(z)  
\right ]_{z=\epsilon} . 
\label{defrho}
\eeqa
As we will show, independently of the mode number $n$, the function $\rho_{\pi^n}$ is 
finite for $m_q \rightarrow 0$. This, like in QCD, allows to predict the behavior of 
$f_{\pi^n}$ close to the chiral limit.

\subsection{Normalization and asymptotic expansion}

It is possible to decouple the system of equations (\ref{EQ1})-(\ref{EQ2}) and 
find an independent equation for the function $\Pi^{n}(z) :=\partial_{z}\pi^{n}(z)$
\beqa
\hspace{-0.5cm}
\left(z\partial_{z}\right)^{2}\Pi^n(z) + A(z) z\partial_{z}\Pi^n(z)
+ B^n(z)\Pi^n(z)=0 \, , 
\label{campoPI}
\eeqa
with $A(z)$ and $B_n(z)$ given by
\beqa
&& A(z) = z \partial_{z}\ln\beta(z)-2 , \\
&& B_n(z) = 1 + z^{2}\left[\partial_{z}^{2}\ln\beta(z)+m_{\pi^n}^{2}-\beta(z) \right].
\eeqa
From (\ref{HGOR}) and (\ref{defrho}) we find that the function $\rho_{\pi^n}$ is 
determined by $\Pi^n$ through the relation 
\begin{equation}
\rho_{\pi^{n}}=-\frac{1}{2m_q g_5 }\left[\frac{\beta (z)}{z}
\Pi^{n}(z)\right]_{z=\epsilon} .
\label{defrho2}
\end{equation}
Expanding $\Pi^n(z)$ in powers of $z$ near the boundary we find the asymptotic 
solution 
\beqa
\Pi^n\left(z\right) &=& C_n \left [ -z+\frac{1}{4}\!\left(m_{\pi^n}^{2}
+\frac{32\pi^{2}\sigma}{3m_{q}}-3m_{q}^{2}\right)\!z^3 + \cdots \right ] 
\nn \\[0.25true cm]
&=:& C_n \, \Pi^n_U (z) , \label{asymptoticsol}
\eeqa
where the dots represent higher powers in $z$ and we have defined $\Pi^n_U(z)$ 
as the  unnormalized function associated with $\Pi^n(z)$. In the numerical 
procedure we will focus on the function $\Pi^n_U(z)$. Note that the solution 
(\ref{asymptoticsol}) naturally satisfies the Dirichlet boundary 
condition (\ref{DirichletBC}).

The constant $C_n$ in Eq.~(\ref{asymptoticsol}) is determined from the normalization 
condition $\Delta_\pi^{nm} = \delta^{nm}$. From Eqs.~(\ref{EQ1}) and (\ref{EQ2}) we 
find that $\Pi^{n}(z)$ obeys normalization condition
\begin{equation}
\int \frac{dz}{z} \, \beta(z)\, \Pi^{n}(z)\Pi^{m}(z) = m_{\pi^{n}}^{2} \delta^{mn},
\label{normalcond}
\end{equation}
which, in turn, leads to
\beqa
C_n = \frac{m_{\pi^n}}{N_{\pi^n}}, 
\label{Cn}
\eeqa
with
\beq
N^2_{\pi^n} = \int \frac{dz}{z} \, \beta(z)\, \left[\Pi^{n}_U(z)\right]^2  . 
\label{normalconst}
\eeq

We end this subsection noticing that, by using the asymptotic expansion (\ref{asymptoticsol}) 
and the definition (\ref{defbeta}), the holographic prescription (\ref{defrho2}) takes the 
simple form
\begin{equation}
\rho_{\pi^{n}}=\frac{g_5 \zeta^{2} }{2 } \frac{ m_{q}  m_{\pi^{n}} }{N_{\pi^n}}  \,.
\label{defrho3}
\end{equation}
This formula will be very useful in the following section. 
 
\section{Numerical results} 
\label{sec:num}

In this section we present our numerical results. We start with the mass spectrum of the pions. 
The spectrum is obtained by solving Eq.~(\ref{campoPI}) for the auxiliary wave function 
$\Pi^n(z) = \partial_z \pi^n(z) $ and imposing the boundary conditions $\pi^n(\epsilon)=0$ 
and $\partial_z \pi^n(z_0) = 0$. We integrate numerically Eq.~(\ref{campoPI}) 
from $z=\epsilon$ to $z=z_0$, using the asymptotic solution (\ref{asymptoticsol}) to extract 
the value for the derivative of $\Pi^n(z)$ at $z = \epsilon$. We use the shooting method, 
which consists in shooting values in the parameter-plane $m_{\pi^n}$ vs $m_q$ until we find 
a solution that satisfies the IR condition $\Pi^n(z_0) = \partial_z \pi^n(z_0)=0$. Due to 
the linearity property of Eq.~(\ref{campoPI}), to get the pion spectrum it is sufficient to 
work with the unnormalized wave function $\Pi_U^n(z)$, defined in Eq.~(\ref{asymptoticsol}). 
This is equivalent to setting $C_n$ to $1$. 

\begin{figure}[htb]
\begin{center}
\includegraphics[scale=0.675]{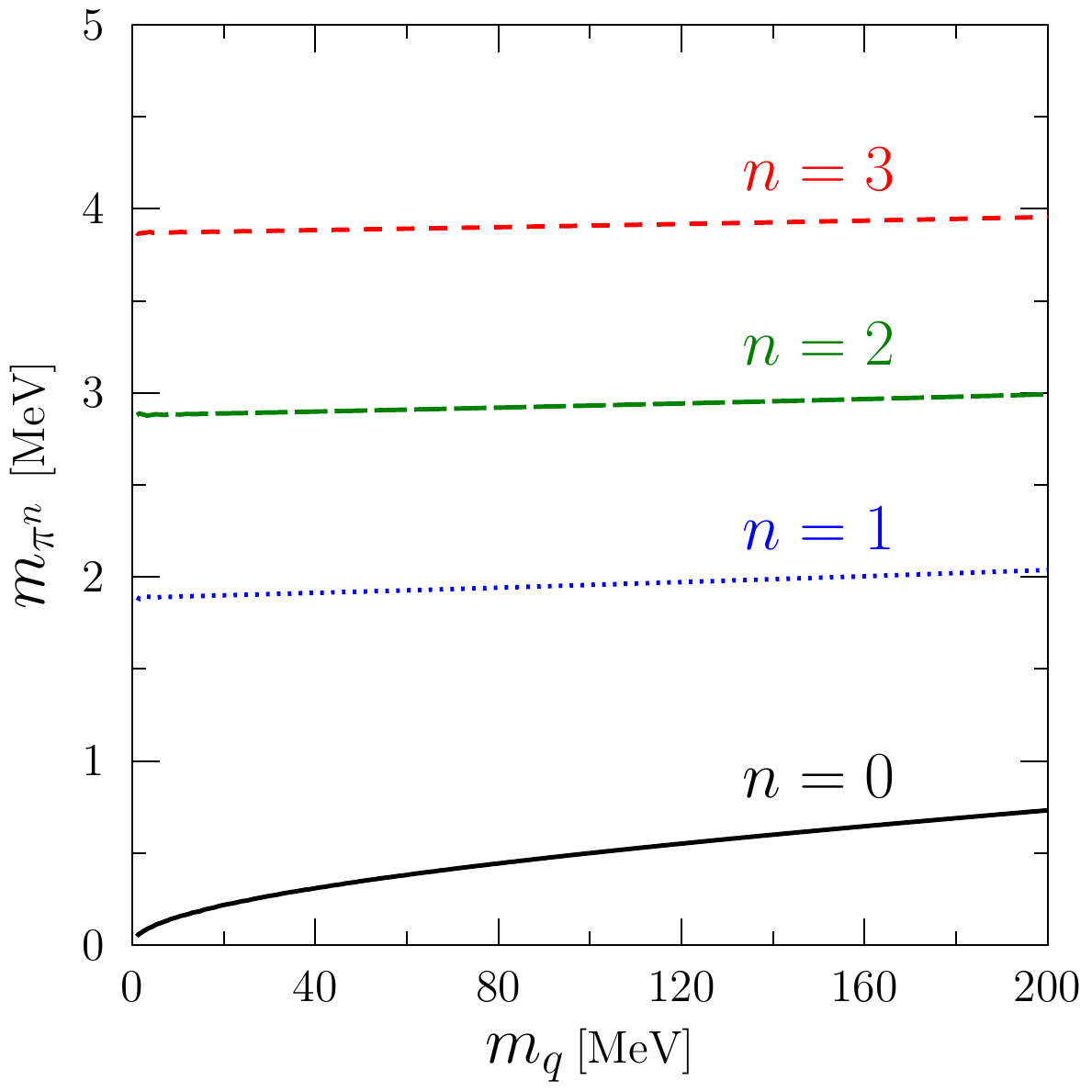}
\end{center}
\vspace{-0.5cm}
\caption{Quark mass dependence of the pion masses.}
\label{masses}
\end{figure}

Parameters are the same used in Refs.~\cite{Erlich:2005qh,Abidin:2009aj}: $m_q = 8.31$~MeV
and $\sigma_q = (213.7 \; {\rm MeV})^3$. They are chosen to fit the ground-state pion mass
$m_{\pi^0} = 139.6$~MeV and leptonic decay constant $f_{\pi^0} = 92.4$~MeV -- the latter
will be discussed further ahead. Fig.~\ref{masses} displays the results for the $m_q$ dependence 
of the pion masses, for the ground state and first three excited states. We have also obtained
solutions for $n > 3$ up to $n = 6$; the $m_q$ dependence of those solutions is similar to 
that shown in Fig.~\ref{masses} for the three lowest excited states. The mass of the ground-state 
pion can be fitted as $m_{\pi^{0}} \sim m_{q}^{1/2}$ near the chiral limit, which is 
consistent with the GOR (\ref{GOR}). On the other hand, the masses of the excited states 
can be fitted with the linear form  $m_{\pi^{n}} = m^0_{\pi^{n}} + a_{n} m_{q}$, 
where $m^0_{\pi^{n}}$ are the corresponding masses in the chiral limit. 
   
It is important to notice that in the hard wall model the squared masses of pion's excited
states grow almost quadratically with the radial excitation number $n$. This a general feature 
of hadron masses in holographic models for QCD that are built in the supergravity approximation, 
where the limit $\lambda \to \infty$ is taken on top of the limit $N_c\to \infty$ . The Regge behavior 
of hadrons (squared masses linear in $n$) is characteristic of strings and an important challenge in 
holography is to reproduce it in terms of an effective 5-d field theory model. The so called soft-wall 
models are very interesting proposals, although the way they may arise from String Theory has not been 
understood. 

\begin{figure}[htb]
\begin{center}
\includegraphics[scale=0.675]{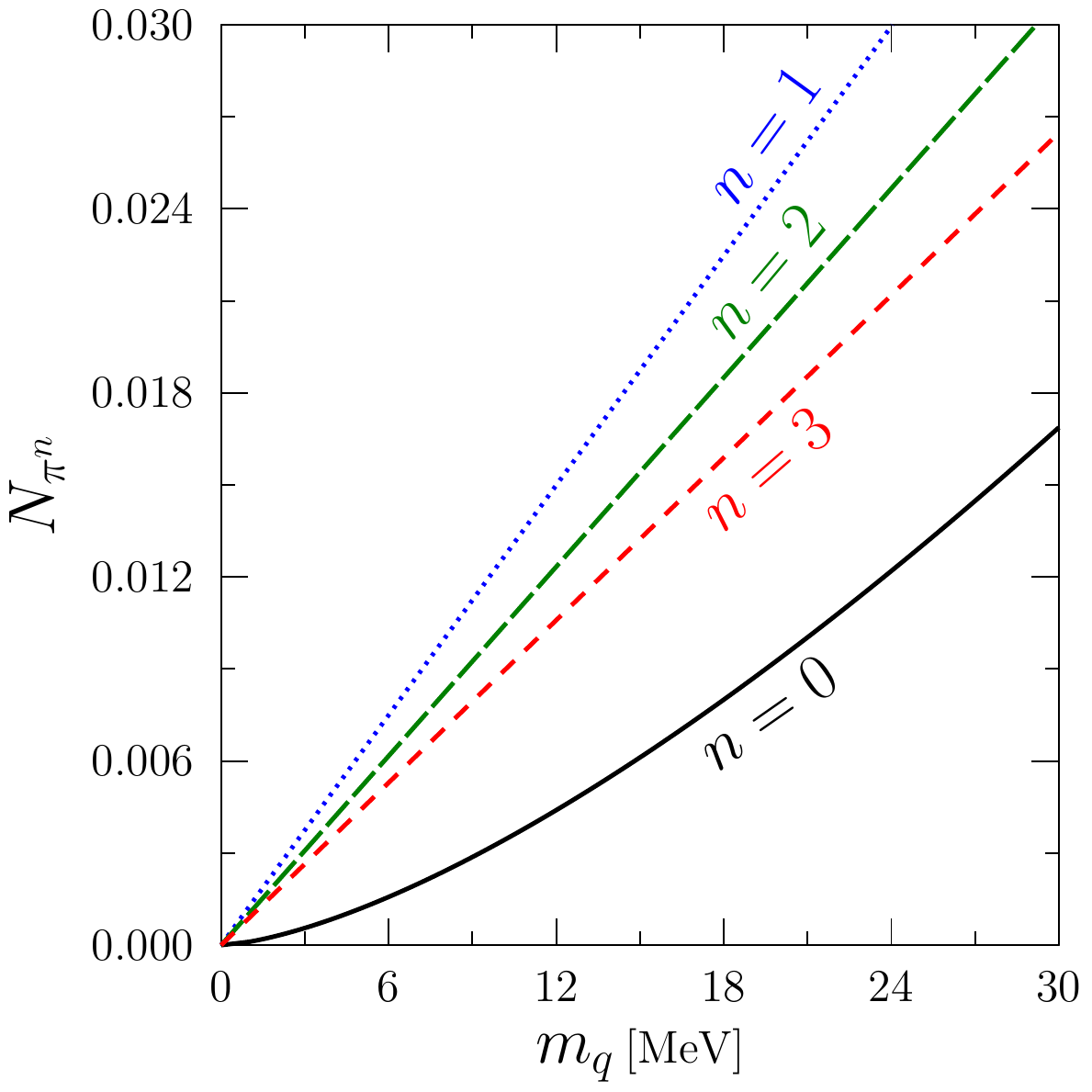} 
\end{center}
\vspace{-0.5cm}
\caption{Quark mass dependence of the normalization constants. }
\label{normal}
\end{figure}

For establishing the finiteness of $\rho_{\pi^n}$ in the chiral limit, we consider 
the $m_q$ dependence of the normalization constant $N_{\pi^n}$, defined in 
Eq.~(\ref{normalconst}). $N_{\pi^n}$ is completely determined from the knowledge of 
the unnormalized wave function $\Pi_U^n(z)$, which enters in the determination
of the mass spectrum. The results are displayed in Fig.~\ref{normal}; the upper panel 
displays the results for the ground state and the lower panel those for the excited states. 
The curve for the ground-state pion can be fitted as $N_{\pi^0} \sim m_q^{3/2}$, while
those for the excited states can be fitted with a linear function 
$N_{\pi^n} \sim  m_q$, $n \ge 1$. As we show next, these different $m_q$ dependences 
of $N_{\pi^n}$ for the ground state and the excited states, when combined with the
different $m_q$ dependence of $m_{\pi^n}$, are responsible for the finiteness of 
$f_{\pi^0}$ and the vanishing of $f_{\pi^n}$ for $n \ge 1$ in the chiral limit.

\begin{figure}[h]
\begin{center}
\includegraphics[scale=0.675]{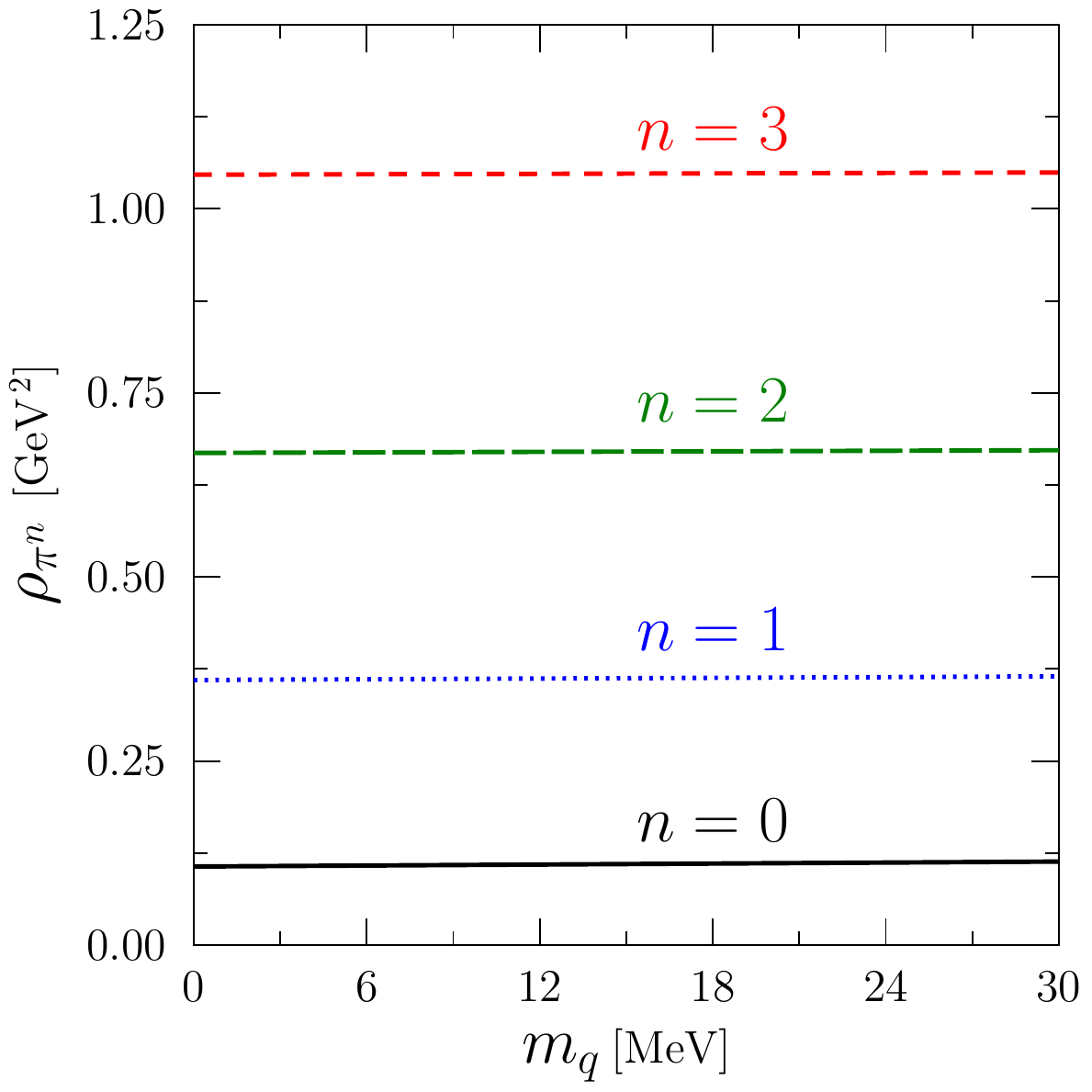}
\end{center}
\vspace{-0.5cm}
\caption{Quark mass dependence of $\rho_{\pi^n}$.}
\label{rhofunctionvsmq}
\end{figure}

As the spectrum and the normalization constant $N_{\pi^n}$ are known, the function 
$\rho_{\pi^n}$ can be readily determined using the formula in Eq.~(\ref{defrho3}). 
Fig.~\ref{rhofunctionvsmq} displays the results. The curves in this figure show that 
$\rho_{\pi^n}$ is finite as $m_q \rightarrow 0$ and remarkably independent of $m_q$ 
for $0 \le m_q \le 30$~MeV for all the values of $n$ investigated. This clearly 
establishes the finiteness of $\rho_{\pi^n}$ at small $m_q$ and, as we discuss next, 
leads to the conclusion that in the chiral limit, $f_{\pi^n} = 0$ for $n \ge 1$. 

\begin{figure}[t]
\begin{center}
\includegraphics[scale=0.675]{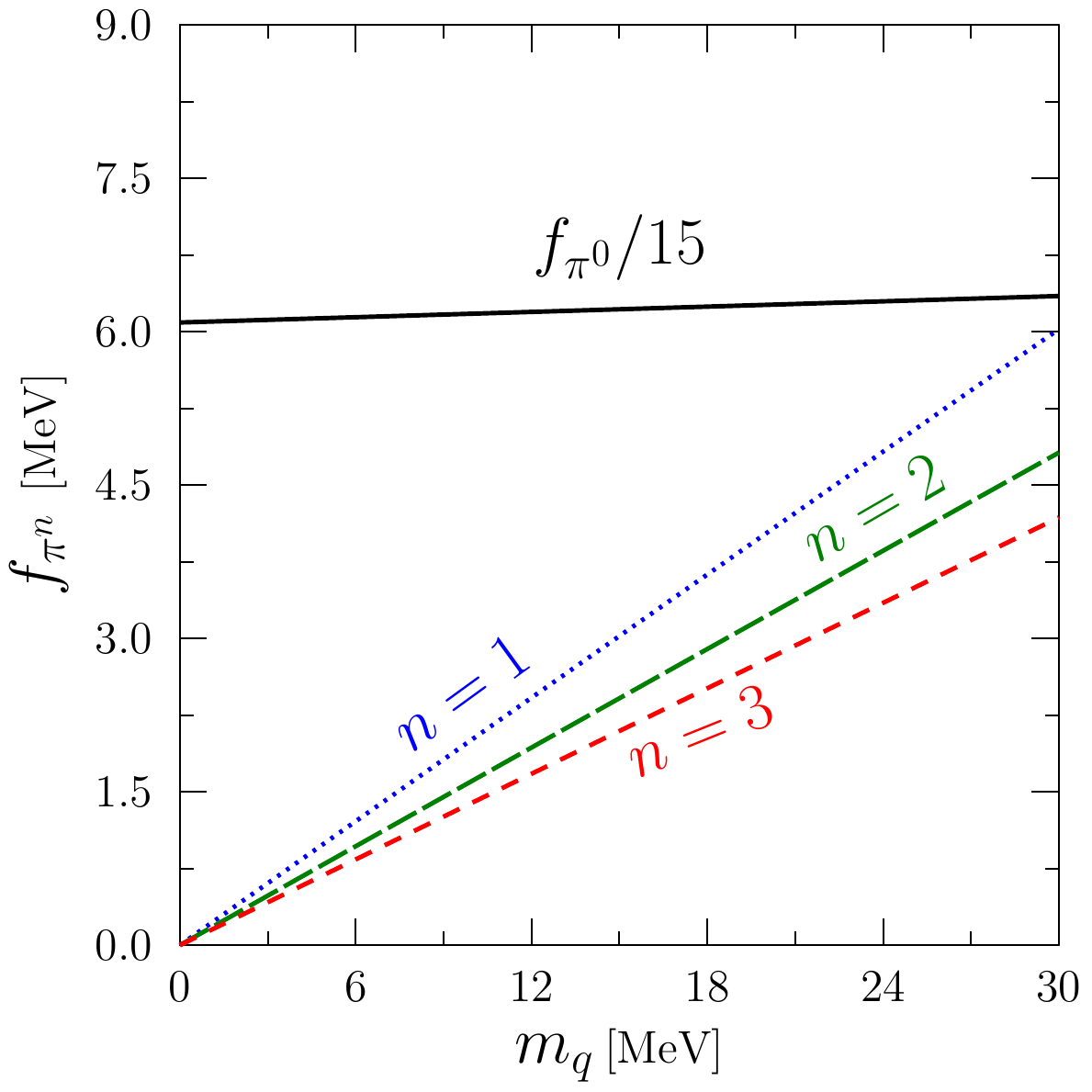}
\end{center}
\vspace{-0.5cm}
\caption{Quark mass dependence of $f_{\pi^n}$. }
\vspace{-0.4cm}
\label{fpivsmq}
\end{figure}

At this stage we are able to present the main result of the present paper, namely the 
behavior of the pion decay constants $f_{\pi^n}$ near the chiral limit. The numerical 
results are displayed in  Fig.~\ref{fpivsmq}. As one can see from the figure, while 
ground-state pion possesses a finite leptonic decay constant $f_{\pi^0}$ the excited 
states have leptonic decay constants $f_{\pi^n}$ that vanish in the chiral limit. 
As mentioned at the beginning of the present section, for $m_q = 8.31$~MeV, the 
ground-state pion decay constant is $f_{\pi^0} \approx 92.4 {\rm MeV}$. Table~\ref{tab:1}
presents results~for~$f_{\pi^n}$.

\begin{table}[htb]
\caption{Leptonic decay constants for the ground-state and the first three 
excited states of the pion.} 
\begin{ruledtabular}
\begin{tabular}{c|cccc}
$n$               &   0   &   1   &  2   &  3  \\ \hline 
$f_{\pi^n}$ (MeV) & 92.4  &  1.68 & 1.34 & 1.16 
\end{tabular}
\end{ruledtabular} 
\label{tab:1}
\end{table}

We also note that the curves for the excited states can be fitted with a linear quark mass 
dependence. Such a linear $m_q$ scaling of $f_{\pi^n}$ for $n \ge 1$ is precisely the one predicted 
in QCD~\cite{Holl:2004fr} through the generalized GOR relationship (\ref{GOR-gen}):
\begin{equation}
f_{\pi^n} = \frac{2 m_q \rho_{\pi^n}}{m^2_{\pi^n}} \sim m_q, \hspace{0.5cm} n \ge 1,
\end{equation}
as $m^2_{\pi^n} \sim (m_q)^0$ and $\rho_{\pi^n} \sim (m_q)^0$.

\begin{figure}[t]
\begin{center}
\includegraphics[scale=0.675]{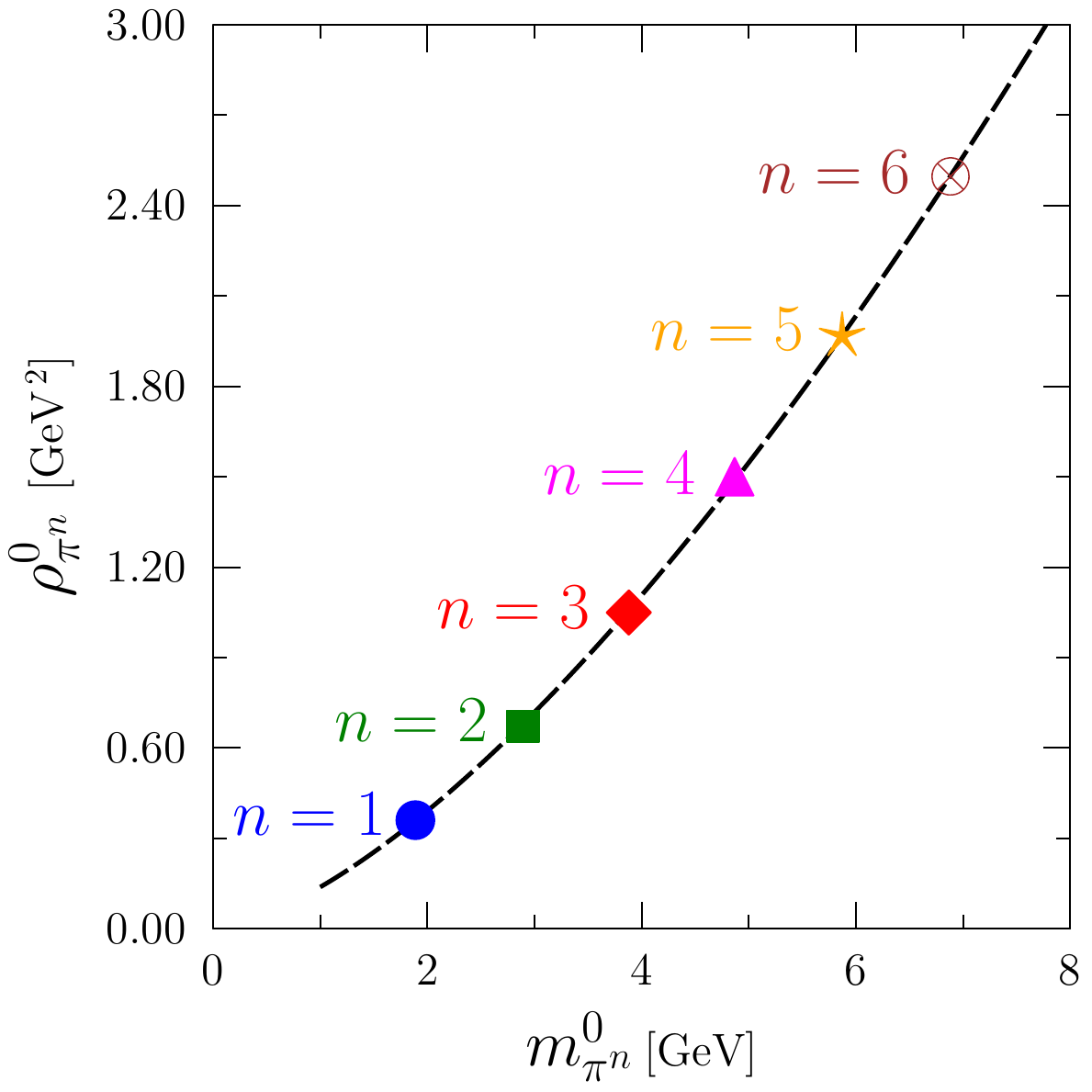}
\end{center}
\vspace{-0.5cm}
\caption{The function $\rho^0_{\pi^n}$, defined in Eq.~(\ref{rho-0}), for the first six 
excited states. The dashed line is a fit to the discrete eigenvalues.}
\label{rhofunctionvsmpi}
\end{figure}

Finally, for completeness we examine $\rho^0_{\pi^n}$, the chiral limit of 
$\rho_{\pi^n}$ defined in Eq.~(\ref{rho-0}), as a function of the pion masses of excited 
states in the chiral limit, $m^0_{\pi^n}$. Note that although the $\rho_{\pi^n}$ are, to a good
precision, $m_q$ independent, the masses $m_{\pi^n}$ are slightly dependent on $m_q$ and this
makes the $m_{\pi^n}$ dependence of $\rho_{\pi^n}$ nontrivial. The results are shown in 
Fig.~\ref{rhofunctionvsmpi}. We found that the six lowest discrete eigenvalues can be 
fitted as 
\begin{equation}
\rho^0_{\pi^n} = \gamma \, \left( m^0_{\pi^n} \right)^{3/2}, \hspace{0.5cm}n \ge 1 ,
\end{equation}
with $\gamma = 4.375 \, {\rm MeV}^{1/2}$. With this, Eq.~(\ref{HGOR}) then implies that the generalized 
GOR relationship takes the form 
\beq
f^0_{\pi^n} := \lim_{m_q \rightarrow 0} f_{\pi^n} = \gamma \, \frac{2 m_q}{\sqrt{m^0_{\pi^n}}}, 
\hspace{0.5cm} n \ge 1.
\eeq

\section{Conclusions and Perspectives} 
\label{sec:concl}

We have investigated the leptonic decay constants of the pion and its excitations in a 
five-dimensional holographic hard wall model for QCD. We have used the model proposed in 
Refs.~\cite{Erlich:2005qh,Da Rold:2005zs} for implementing dynamical chiral symmetry breaking 
and introduced a direct way of calculating the decay constants via the definition of holographic 
currents. We have proved numerically that the leptonic decay constants of the excited states of 
the pion vanish in the chiral limit. In addition, we have shown that these results follow from 
the generalized GOR relationship Eq.~(\ref{GOR-gen}), whose counterpart in QCD was first
derived in QCD in Ref.~\cite{Holl:2004fr}.  

Our results for the vanishing of the leptonic decay constants of pion's excited states in
the chiral limit, besides being in agreement with QCD, might shed light on the failure of 
light-front holography (LFH) in reproducing such results. A key feature of the 
approach we followed in the present paper is the generalized GOR relationship of 
Eq.~(\ref{GOR-gen}). The generalized GOR relationship is a direct consequence of the 
dynamical breaking of chiral symmetry implemented by the action of Ref.~(\ref{Erlichaction}) 
via a scalar field $X(x,z)$ that has a negative mass squared, which leads to the extended 
PCAC relation given in Eq.~(\ref{ext-PCAC}). On the other hand, as mentioned in the Introduction, 
the way chiral symmetry is treated in LFH is nonstandard, as the vanishing of the pion mass in the 
chiral limit is not a result of the dynamical breaking of the symmetry~\cite{Brodsky:2013ar}. 
In LFH the vanishing of the pion mass when $m_q = 0$ follows from the precise cancellation of 
the light-front kinetic energy and light-front potential energy terms for the quadratic confinement 
potential in a Schr\"odinger-like equation~\cite{Brodsky:2013ar}. Although similar cancellations
occur in chiral models of QCD in Coulomb gauge~\cite{{Amer:1983qa},{Adler:1984ri},{Bicudo:1989sh},
{Bicudo:1991kz},{Szczepaniak:2001rg},{Alkofer:2005ug},{Pak:2011wu},{Fontoura:2012mz}}, there is however
one crucial aspect in the calculation of the pion leptonic decay constants in LFH~\cite{Branz:2010ub} 
that might not capture the full chiral dynamics of the pion bound state, namely the truncation of 
the Fock-space decomposition of pion's light-front wavefunction to its lowest quark-antiquark valence 
component. In this sense, it would be interesting to find means to include higher Fock-space components
in pion's wave function in LFH.  

Notwithstanding our results are derived in a hard wall model of holographic QCD, we believe that
the vanishing of the leptonic decay constants of pion's excited states in the chiral limit 
will happen in any holographic model that implements dynamical chiral symmetry breaking and 
reproduces the generalized GOR relationship. This and other aspects of the problem investigated here
should be investigated with soft-wall models. Besides improving on the mass spectrum, such a model
make closer contact with LFH.

\section*{Acknowledgements}

The work of A.B. has been supported by the Portuguese agency Funda\c{c}\~ao para a 
Ci\^encia e a Tecnologia - FCT through the fellowship SFRH/BI/52142/2013 and the Brazilian 
agency Coordena\c{c}\~ao de Apoio ao Pessoal de Nivel Superior - CAPES, through the fellowship 
BEX 8051/14-3. The work of G.K. was supported in part by the Brazilian agencies Conselho Nacional 
de Desenvolvimento  Cient\'{\i}fico e Tecnol\'ogico- CNPq, Grant No. 305894/2009-9, and 
Funda\c{c}\~ao de Amparo \`a Pesquisa do Estado de S\~ao Paulo - FAPESP, Grant No. 2013/01907-0. 
C.M. was supported by a doctoral fellowship of the Brazilian agency Coordena\c{c}\~ao de Apoio 
ao Pessoal de Nivel Superior - CAPES.


\end{document}